# THE CHEMICAL ORIGIN OF SEY AT TECHNICAL SURFACES

R. Larciprete, ISC-CNR, Roma and LNF-INFN, Frascati (Italy)
D. R. Grosso, M. Commisso LNF-INFN Frascati (Italy)
R. Flammini, IMIP- CNR, Montelibretti and LNF-INFN, Frascati (Italy)
R. Cimino, LNF-INFN, Frascati (Italy)

*Abstract*

The secondary emission yield (SEY) properties of co-laminated Cu samples for LHC beam screens are correlated to the surface chemical composition determined by X-ray photoelectron spectroscopy. The surface of the "as received" samples is characterized by the presence of significant quantities of contaminating adsorbates and by the maximum of the SEY curve ($\delta_{max}$) being as high as 2.2. After extended electron scrubbing at kinetic energy of 10 and 500 eV, the $\delta_{max}$ value drops to the ultimate values of 1.35 and 1.1, respectively. In both cases the surface oxidized phases are significantly reduced, whereas only in the sample scrubbed at 500 eV the formation of a graphitic-like C layer is observed.

We find that the electron scrubbing of technical Cu surfaces can be described as occurring in two steps, where the *first step* consists in the electron induced desorption of weakly bound contaminants that occurs indifferently at 10 and at 500 eV and corresponds to a partial decrease of $\delta_{max}$, and the *second step*, activated by more energetic electrons and becoming evident at high doses, which increases the number of graphitic-like C-C bonds via the dissociation of adsorbates already contaminating the "as received" surface or accumulating on this surface during irradiation. Our results demonstrate how the kinetic energy of impinging electrons is a crucial parameter when conditioning technical surfaces of Cu and other metals by means of electron induced chemical processing.

## INTRODUCTION

A wide range of applications [1-3] use or are dependent on the capability of a given material to emit electrons after electron bombardment. This quantity, called secondary electron yield (SEY), is defined as the ratio of the number of emitted (or secondary) electrons to the number of incident primary electrons [4], and is commonly denoted by $\delta$. The SEY $\delta_{max}$ curves, which are characterized by the behavior at low energy [5] and by the asymptotic value at high energy of the incident electrons, for many purposes can be schematically described by their maximum value ($\delta_{max}$) and the energy at which it occurs ($E_{max}$). Our experiments are performed in the context of particle accelerator research, since, when intense and positively charged beams are circulating in vacuum chambers of small transverse dimensions, may interact with low energy electrons also present in the vacuum chamber loosing the desired properties. The low energy electrons, produced either by synchrotron radiation hitting the accelerator walls [6,7] or by direct ionization of residual gases, might undergo a rapid multiplication driven by the actual SEY properties of the wall surface. In fact the seeding primary electrons "see" the circulating beam and are accelerated in a complex dynamics (studied in details in different simulation codes developed to this purpose) [8-12] and hit the vacuum wall. The secondary electrons are produced and a multiplication, resonant with the beam time structure, may occur if the accelerator wall surface possesses a SEY larger than unity. This can cause a sudden increase of the number of electrons in the accelerator, inducing detrimental effects on beam quality as well as rapid vacuum pressure rises resulting in beam loss. This phenomenon is called electron cloud (EC) build-up, and has been recognized as a problem in positron/proton rings like DAFNE, B (Beauty) factories, PEP-II, KEKB [13-17] and LHC among others.

A mean to mitigate this problem is to exploit the conditioning or scrubbing effect that the prolonged electron irradiation has on the chemical state of the wall surface and that often coincides with a significant reduction of the SEY [13-16]. LHC, for instance, bases its ability to run at operation conditions on a drastic reduction of the initially high SEY ($\delta_{max} \sim 2.2$) of the Cu surface seeing the beam in the cryogenic dipoles to a much lower value ($\delta_{max} \sim 1.3$) after a certain electron dose. Electron scrubbing is considered then necessary to reach nominal operation [13-15, 18].

Scope of this study is the detailed comprehension of the chemistry variations induced by electron irradiation on technical surfaces, i.e. samples representative of the accelerator walls, exposed to air and not treated with specific cleaning procedures in vacuum. Recent studies have demonstrated that the beneficial effect of electron beam scrubbing on these surfaces in some cases coincides with the formation of a graphitic surface film [19,20]. Since the SEY of graphite, and in general of carbon based materials is lower than that of air exposed metals, the presence of the C thin films reduces the effective SEY of the surface [21]. Graphitic film growth occurs because, in general, the technical surfaces are covered by C containing contaminants that once exposed to the electron flux tend to decompose and partly rearrange in graphitic assemblies [22,23]. The occurrence of material transformation at the atomic level induced by electron irradiation has been often reported in the case of thin films and nanostructures [24,25]. In particular, the graphitization of carbonaceous films is a frequent process and relies on the higher stability of graphitic lattice at or below ambient pressure over the other possible C allotropic structures. The electron-induced chemical reactions at the basis of contaminant graphitization lead to the dissociation of C-H [26,27] and C-O bonds and to the formation of volatile compounds that desorb from the surface. In parallel C-C bonds reorganize from the open-chain geometry, typical of aliphatic hydrocarbons, to form domains with the honeycomb arrangement characteristic of the graphitic materials, due to the transition of the C atoms from the $sp^3$ to $sp^2$ hybridization state. Moreover, since the incident electrons are emitted by a hot filament, this, if not properly degassed, could contribute to a local increase of C-containing contaminants. In addition to that, the electron beam might also induce the deposition of a thin graphitic layer by dissociating C-containing gas phase molecules present in the residual pressure of the vacuum chamber [28].

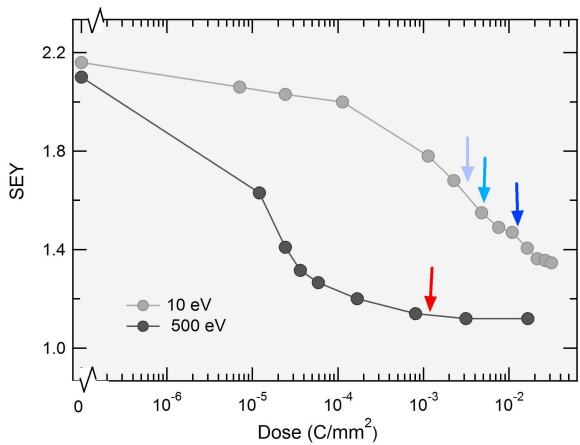

Figure 1 $\delta_{max}$ values measured on co-laminated Cu samples for LHC beam screen as a function of the electron dose at $E_p$=500 eV and 10 eV [20]. The arrows indicate the doses used in this experiment.

This process, that is certainly more relevant in low vacuum environments, might occur even in ultra high vacuum regimes ($10^{-9}$-$10^{-10}$ mbar) due to the dissociation of molecules such as CO and $CO_2$ that are usual components of the residual gas. In fact, the growth of thin carbon layers is routinely observed on surfaces exposed to high energy radiation as in electron microscopy [29], extreme ultraviolet lithography [3,30] or synchrotron radiation beamlines [31]. Both the graphitization of the pre-existing contaminating layer and the growth of a graphitic film due to the cracking of the residual gas molecules occur with a different efficiency depending on the kinetic energy of the electrons used to scrub the surface.

Recently the effect of the kinetic energy of the scrubbing electrons on the SEY has been investigated in the case of co-laminated Cu for LHC beam screen which, when characterized "as received" shows a $\delta_{max}$ of 2.2 [20]. At each kinetic energy of the primary beam $E_p$ between 10 and 500 eV, electron scrubbing was found to lower the SEY, with $\delta_{max}$ decreasing asymptotically down to an ultimate minimum value, which, for kinetic energy between 50 and 500 eV is 1.1, whereas for kinetic energy of 10 eV remains around 1.35 [20]. The stability of the $\delta_{max}$ values after further irradiation indicates that the samples are in each case "fully scrubbed" at the corresponding energy. This is shown in Fig.1 for the $\delta_{max}$ curves taken at $E_p$ = 10 and 500 eV. As a consequence the majority of the electrons forming the e-cloud in the LHC, having energy below 20 eV [20], do not contribute in lowering $\delta_{max}$ below the value of 1.3 desired for machine stability at design operation [13-15, 18]. The relevance of this issue for accelerator wall conditioning motivates deep investigations of the effects of the electron kinetic energy on the surface chemistry of technical metal surfaces.

In this study the SEY properties of co-laminated Cu samples for LHC beam screen scrubbed at 10 and 500 eV were correlated to the surface chemical composition determined by X-ray photoelectron spectroscopy (XPS). Our results show that electron scrubbing at 10 eV efficiently removes many contaminating species from the sample surface diminishing significantly the oxygen content, but fails to induce a substantial graphitization. In contrast, the formation of a graphitic C layer is clearly observed on the surface scrubbed at $E_p$= 500 eV, whose SEY is satisfactorily mitigated.

## EXPERIMENTAL

XPS and SEY measurements were performed at the Material Science Laboratory of the INFN-LNF in Frascati (RM). The experimental apparatus is described in detail elsewhere [6]. Briefly the UHV system (base pressure $2\times10^{-10}$ mbar) includes a μ-metal chamber dedicated to SEY measurements and XPS analysis and a preparation chamber. The SEY (δ), i. e. the ratio of the number of electrons leaving the sample surface (*Is*) to the number of incident electrons (*Ip*) per unit area, is determined experimentally by measuring *Ip* and the total sample current $I_T=Ip - I_S$ so that δ=$1-I_T /I_P$. For the SEY measurements, the electron beam was set to be smaller than 0.25 mm² in transverse cross-sectional area at the sample surface. To measure the current of the impinging primary electrons, a negative bias voltage (-75 V) was applied to the sample. The SEY measurements and electron irradiation were performed at normal incidence, by using electron beam currents of a few nA (to induce minimal "scrubbing" during data acquisition) and ~1-5 μA, respectively. In order to take XPS spectra in the electron irradiated regions the electron beam was scanned to scrub a 3×3 mm² area, and therefore the doses delivered to the sample in this experiment were lower than those reached when irradiating a fixed point [20]. The SEY was found to fluctuate by 5% at most. XPS spectra were acquired by exciting the sample with non-monochromatic MgKα photons (hν=1253.6 eV) and by detecting the photoelectrons in normal emission geometry by means of an hemispherical electron analyzer. The field of view of the electron analyzer was smaller than 1.5 mm². The binding energies (BE) are referred to the Fermi level measured on the sample.

## RESULTS AND DISCUSSION

Figure 2a shows the SEY curve measured on the "as received" sample (trace A) that exhibits a $\delta_{max}$ value of 2.1. The XPS spectrum taken on this surface (see Fig.2c) shows the Cu spectral features but also reveals the presence of C and O indicated by the C1s and O1s peaks at around 285 and 531 eV respectively, due to surface contaminants after the prolonged permanence in air. The C1s core level spectrum, shown in Figs.2d, consists of a main structure peaked at 284.6 eV and a weaker peak centered around 288.1 V. The first peak can be related to the presence of C-C and C-H bonds, with the C atoms having on the average a hybridization state intermediate between $sp^3$ and $sp^2$, that are characterized by typical BE values of ~284.3 and ~285.1 eV, respectively [32,33], as indicated by the arrows in Fig.2d. This chemical arrangement derive from the different compounds forming the adsorbed layer. The high BE tail of the main C1s peak as well as the weaker peak at 288.1 eV are indicative for the presence of C-OH, C-O-C and C=O bonds [34].

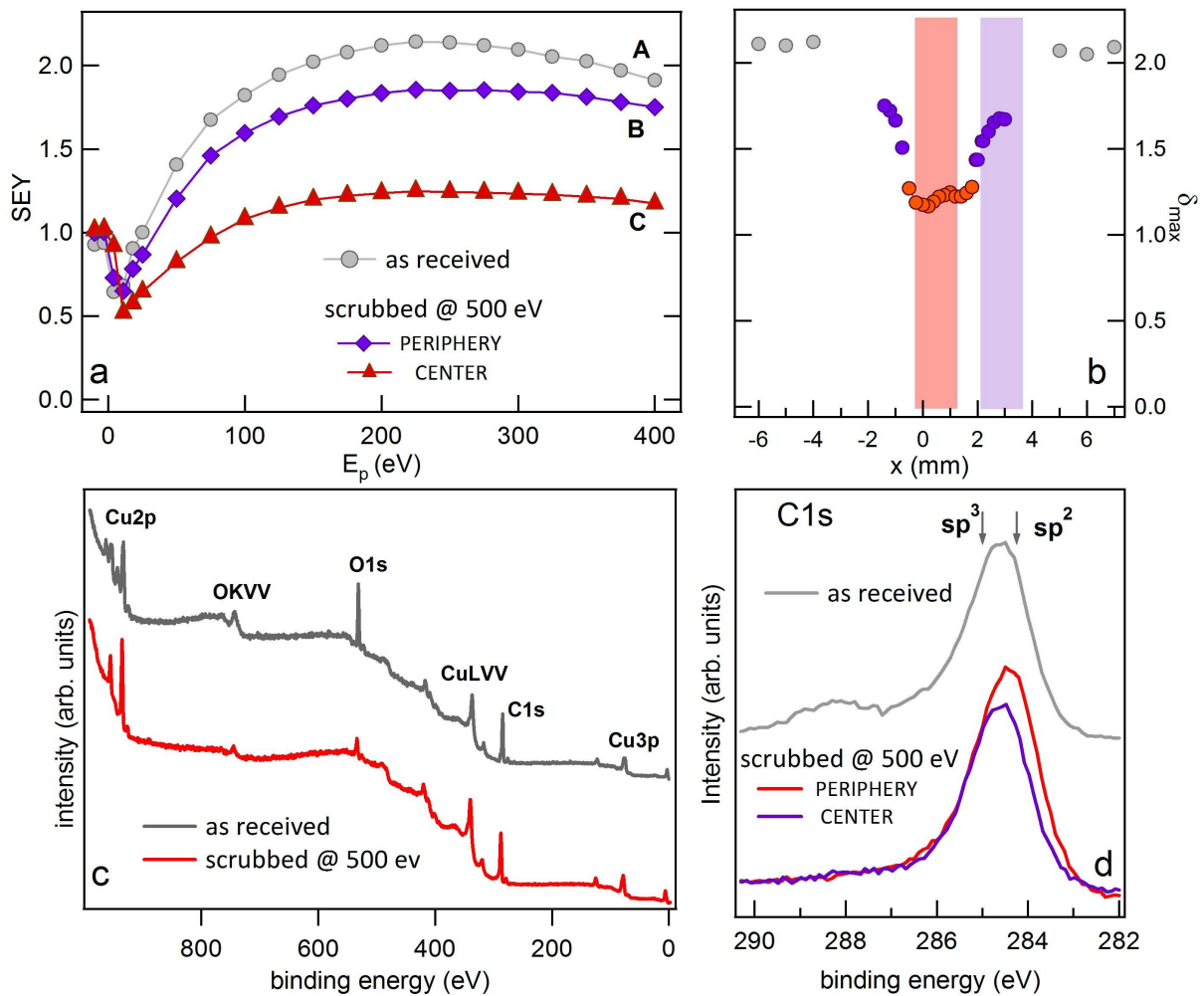

Figure 2: a) SEY curves measured on the "as received" co-laminated Cu sample for LHC beam screen (grey, circles) and in the middle (red, triangles) and in the periphery (violet, diamonds) of the sample area (3×3 mm²) scrubbed with 500 eV electrons (Q=1.2×10⁻³ C/mm²); b) $\delta_{max}$ values measured along a line crossing the scrubbed region; c) XPS spectra measured on the "as received" sample (upper curve) and in the middle of the scrubbed region (lower curve); d) C1s core level spectra measured on the "as received" sample; in the periphery and in the center of the scrubbed region.

This sample was irradiated at a primary electron energy $E_p$=500 eV over an area of 3×3 mm$^2$ obtained by scanning the electron beam up to a total dose of Q=1.2×10$^{-3}$ C/mm$^2$. According to the Fig.1 this corresponds to have the surface almost "fully scrubbed". The effect of electron scrubbing was studied by comparing the chemical composition and the secondary electron emission measured outside and inside the irradiated area. After electron conditioning the SEY curve measured in the center of the irradiated area shows a $\delta_{max}$ value of 1.2 (see trace C in Fig.2a). This value is almost constant over a length of 2.3 mm along a line crossing the electron beam spot (red area in Fig.2b), indicating an homogeneous scrubbing effect over the region. Higher values are measured in the periphery of the scrubbed region (purple area in Fig.2b), in correspondence of the tails of the electron beam, where the delivered electron dose is lower. The SEY curve measured in the periphery of the scrubbed region (trace B in Fig.2a) shows a $\delta_{max}$ of 1.8. However, this value is strongly dependent on the exact position of the sampled point, as the highly sloping $\delta_{max}$ curve shown in Fig.2b indicates. On the other hand, far away from the irradiated region (grey circles in Fig.2b) the sample maintains the $\delta_{max}$ values typical of the "as received" surface. The variation of the secondary emission corresponds to significant modifications of the surface chemical composition. The XPS spectrum measured in the center of the irradiated region is shown in Fig.2c whereas Fig.2d compares the C1s spectra taken in the center and in the periphery of this area. In the periphery region (violet curves), that is, in the area scrubbed at a lower electron dose, the C1s spectrum has lost the C-O component. Consistently the intensity of the O1s spectrum (not shown) has decreased substantially. These chemical modifications are likely due to the dissociation of Cu-O, C-H [26,27] and C-O bonds and to the recombination of volatile molecules as $O_2$ and $H_2O$ that easily desorbed under the action of the impinging electrons. In this reaction, a possible role of secondary electrons coming from the bulk of the sample cannot be excluded [35]. The loss of O containing molecules reduces the oxidizing components in the contaminated surface and results in a SEY decrease [36].

In the center of the scrubbed area the amount of O is even lower and the C peak has shifted to lower BE. [20] This means that, in addition to the reactions occurring at the periphery of the beam spot, here the impinging electrons have also converted the C hybridization from sp$^3$ into sp$^2$. Such effect usually is accompanied by a decrease of the SEY of technical surfaces [19,20]. Moreover in the scrubbed area the C1s intensity which is ~ 20% higher than in the periphery, hints at the occurrence of electron beam induced deposition of graphitic C. The additional C layer growth originates from the dissociation of residual gas molecules present in the UHV chamber or even released by the hot e$^-$ beam filament, typically CO and $CO_2$, that adsorb on the sample surface where they are cracked by the impinging 500 eV electrons. After the dissociation, the O atoms desorb as $O_2$ whereas the C atoms bind to each other and condense in graphitic-like organized network.

The effect of the kinetic energy of the impinging electrons on the conditioning of the LHC sample was investigated by performing a similar irradiation experiment on a second "as received" sample at $E_p$ of 10 eV over a 3x3 mm$^2$ area. Fig.3 shows the SEY curves measured after electron doses of 3.2×10$^{-3}$ and 4.8×10$^{-3}$ C/mm$^2$ that exhibit $\delta_{max}$ values of 1.64 and 1.54 whereas a final value of 1.46 is reached after a dose of 1.1×10$^{-2}$ C/mm$^2$, which is still lower than that required to fully scrub the sample at 10 eV, as shown by the curve plotted in Fig.1.

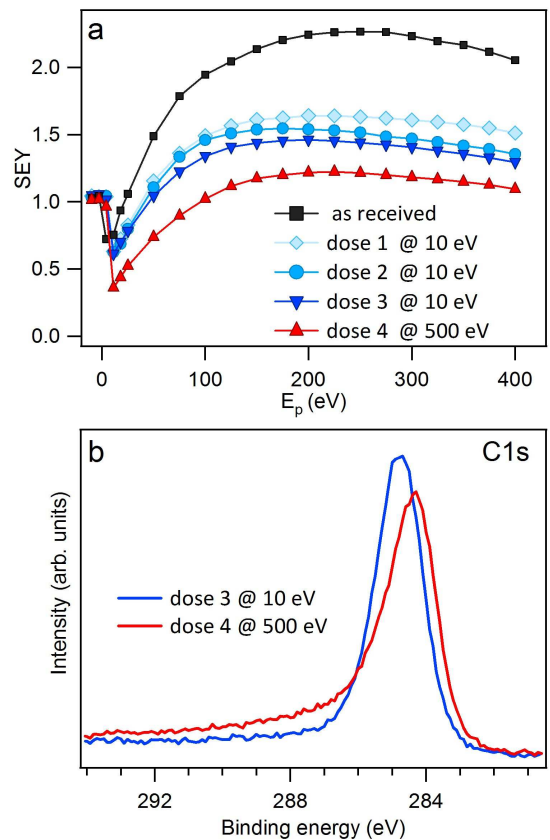

Figure 3: a) SEY curve measured on the co-laminated Cu sample for LHC beam screen "as received" and in the middle of the area (3×3 mm$^2$) scrubbed with 10 eV electrons (dose 1: Q=3.2×10$^{-3}$ C/mm$^2$; dose 2: 4.8×10$^{-3}$ C/mm$^2$; dose 3: 1.1×10$^{-2}$ C/mm$^2$), and afterwards scrubbed with 500 eV electrons (dose 4: Q=1.2×10$^{-3}$ C/mm$^2$); b) C1s core level spectra taken on the sample surface scrubbed at 10 eV and subsequently fully scrubbed at 500 eV...

Correspondingly the C1s spectrum shows a single symmetric peak centered at BE of 284.7 eV as on the "as received" sample (see Fig.2d), whereas the peak at

BE of 288.1 eV, that before irradiation was indicative of C-H, C-O and C=O bonds (see Fig.1d), has disappeared. This shows that the prolonged electron scrubbing at $E_p$ of 10 eV has successfully removed the oxidized components resulting from O containing contaminants, but is not effective in converting the hybridization state of the C atoms to form graphitic domains. The decrease of $\delta_{max}$ from 2.1 to 1.46 has then to be related to the reduction of the surface contaminants after the desorption of oxygen-carrying species.

To confirm that the key factor in fully reducing the SEY is the kinetic energy of the impinging electron, this sample was subsequently scrubbed at $E_p$=500 eV up to a dose of $1.2\times10^{-3}$ C/mm$^2$. Then, an efficient graphitization was obtained as indicated by the C1s line shape, that after such additional scrubbing exhibits the asymmetric profile peaked at 284.3 typical of graphitic carbon (Fig.3b). In this particular case the growth of additional C under the action of the beam is marginal as shown by the comparable C1s intensities measured before and after the scrubbing at 500 eV. In agreement with the behavior observed before, the SEY curve measured on the graphitized surface shows a $\delta_{max}$ value of 1.2, confirming the beneficial effect of ultra thin graphitic-like C films on the secondary emission properties of copper technical surfaces.

By combining the results obtained at $E_p$ of 500 and 10 eV it is possible to describe the electron scrubbing of technical Cu surfaces as occurring in two steps, where the *first step* consists in the electron induced desorption of weakly bound contaminants that occurs indifferently at 10 and at 500 eV and corresponds to a partial decrease of $\delta_{max}$, and the *second step*, activated by more energetic electrons and becoming evident at high doses, which increases the number of graphitic-like C-C bonds via the dissociation of adsorbates already contaminating the "as received" surface or accumulating on this surface during irradiation.

## CONCLUSIONS

We have shown that the SEY of co-laminated Cu sample for LHC beam screen can be decreased by electron scrubbing. However, lowering the initial $\delta_{max}$ of 2.1 to values below 1.4 requires the formation of a graphitic film. This occurs via electron beam induced reactions in the C-containing contaminating layer covering the "as received" sample, and/or via electron beam induced dissociation of adsorbates coming from the residual gas of the vacuum chamber or released by the e$^-$ gun filament (typically CO and $CO_2$), whose fragments partly desorb and partly organize in graphitic domains. Due to this evidence some concern arises with respect to the comparison of data taken at different base pressure and by using differently degassed electron beam filaments. Undoubtedly more systematic studies are required to fully understand the processes leading to surface conditioning.

We confirm the mitigating effect of thin graphitic films on the surface SEY, and demonstrate the limited scrubbing effectiveness of the low kinetic energy electrons. As a matter of fact neither the low kinetic energy impinging electrons ($E_p$=10 eV) nor the low energy secondary electrons coming from the bulk of the sample are efficient towards surface graphitization, but the interaction with energetic electrons seems to be indispensable to convert the adsorbed C atoms into a graphitic-like network. These results, having a direct relevance for LHC, might also widen the general perspective of accelerator wall conditioning and may be of interest to the much wider community studying the SEY surface properties in various fields of research.

## ACKNOWLEDGMENT

This work has been supported by INFN- NTA and Group V. The assistance of the technical team of DAFNE-L is deeply acknowledged.